\documentclass{iucrjournals}

\usepackage{amsmath}   
\usepackage{amssymb}  
\DeclareMathOperator*{\argmin}{arg\,min}

\usepackage{amsthm}    
\usepackage{mathtools}
\usepackage{subcaption}

\usepackage[T1]{fontenc}
\usepackage{lmodern}
\usepackage{microtype} 
\usepackage{siunitx}
\usepackage{hyperref} 
\usepackage[nameinlink,capitalise]{cleveref}   
\theoremstyle{plain}

\usepackage{bm}        
\usepackage{algorithm}
\usepackage{algpseudocode}  
\nolinenumbers

\title{Robust Indexing for Challenging Serial X-ray Diffraction Patterns}

\author[a,b]{Marc M Nasser\IUCrCemaillink{mnasser@stanford.edu}\IUCrOrcidlink{0009-0004-9251-6094}}%
\author[b]{Frédéric Poitevin\IUCrEmaillink{frederic.poitevin@stanford.edu}\IUCrOrcidlink{0000-0002-3181-8652}}%
\author[b]{Kevin M Dalton\IUCrEmaillink{kmdalton@slac.stanford.edu}\IUCrOrcidlink{0000-0001-9396-085X}}%

\affil[a]{Electrical Engineering, Stanford University}
\affil[b]{LCLS Data Systems, SLAC National Accelerator Laboratory}

\begin{document} 
\maketitle 

\begin{abstract}
Serial crystallography experiments routinely produce thousands of diffraction patterns from crystals in random orientations. To turn this stream of images into a usable dataset, each pattern must be indexed (having its crystal orientation recovered from a small set of measured reflections) before integration and merging can proceed. In practice, diffraction patterns can contain only a small number of reliable peaks, are contaminated by background and spuriously detected reflections, or arise from crystals with highly skewed unit cells. These factors make indexing unstable precisely in the small-$N$ regime.

We introduce a robust indexing algorithm tailored to this setting. We formulate indexing as a symmetry-aware lattice decoding problem and design a loss that explicitly incorporates lattice symmetries while trimming outlier peaks that are inconsistent with any plausible orientation. We combine this objective with a reciprocal-space basis reparameterization that stabilizes decoding for skewed or poorly conditioned lattices, and we develop a dedicated small-$N$ objective mode that couples refined peak scoring with a method to recover orientations from very few reflections. The resulting method can be implemented in a memory-efficient way and used as a robust indexer.

We evaluate our approach on three protein datasets from the Coherent X-ray Imaging Data Bank collected at XFEL facilities, using identical preprocessing and unit-cell information across methods. Across all datasets, our algorithm matches or outperforms established indexers such as XGANDALF and TORO, with particularly large gains for patterns with few indexed peaks and for crystals with skewed unit cells. While slower, our algorithm is extremely memory-efficient, and its structure readily enables high-degree parallelization on CPUs or larger batch sizes on GPUs. Our results show that carefully exploiting lattice structure, symmetry, and small-$N$–aware search yields substantial improvements in indexing robustness.
\end{abstract}


%------------------------------------------------------------------------------
\section{Introduction}

\subsection{Motivation and Background}
Serial crystallography has become a central approach for determining macromolecular structures at both X-ray free-electron laser (XFEL) facilities.
In serial experiments, a large number of diffraction patterns are collected from many microcrystals delivered in a continuous stream. Each pattern typically samples a different random crystal orientation and contains only a subset of all possible reflections. To reconstruct a full reciprocal-space dataset, each frame must be \emph{indexed}: one must determine the crystal orientation and assign Miller indices to the observed peaks. Indexing accuracy directly affects downstream integration, merging, and phasing, and is therefore a crucial component of modern serial crystallography pipelines.

As such, it is important to robustly and efficiently index large numbers of sparsely populated, noisy diffraction patterns. Existing indexers have made substantial progress towards this goal, particularly in computational speed, yet a nontrivial fraction of frames in many datasets remain unindexed, reducing completeness and the quality of the final merged structures.

In this work we focus on the common scenario where approximate unit-cell parameters are known and supplied to the indexer. This matches the standard usage in serial crystallography pipelines such as CrystFEL/indexamajig \cite{crystfel}, where a unit-cell file is typically provided alongside the detector geometry and other experimental parameters~\cite{crystfel}.

\subsection{Indexing Challenges in Serial Crystallography}
Several factors make indexing particularly challenging in serial crystallography:

\begin{itemize}
    \item \textbf{Sparse peak sets.} Many diffraction patterns contain only a small number of measured reflections, for example from weakly diffracting samples or partially illuminated crystals. In this low-$N$ regime, existing indexers typically experience a marked decline in performance compared to patterns with larger numbers of reflections, as their methods are sensitive to per-peak assignment errors.

    \item \textbf{High noise and false peaks.} Background scattering, detector artifacts, and peak-finding errors introduce outliers that can mislead least-squares or grid-based search methods, especially since loss landscape is highly non-convex.

    \item \textbf{Skewed or ill-conditioned reciprocal bases.} Conventional unit-cell bases may lead to poorly conditioned integer decoding, especially in low-symmetry systems or poorly chosen primitive cells.

    \item \textbf{Throughput and deployment constraints.} High repetition-rate XFELs and fast serial synchrotron beamlines generate data at rates that challenge both online (real-time) and offline analysis pipelines. Indexers must balance robustness against computational cost, memory usage, and latency.
\end{itemize}

These difficulties motivate the development of indexing methods that (i) remain robust for small-$N$ patterns, (ii) handle noise and outliers effectively, (iii) leverage lattice symmetries and basis reduction for stable integer decoding, and (iv) explore the orientation space efficiently, balancing speed and memory, making them usable across both XFEL and synchrotron serial crystallography experiments.

\subsection{Related Work}
Several indexing algorithms have been developed for serial crystallography, with PinkIndexer~\cite{gevorkov2020pinkindexer} and XGANDALF~\cite{gevorkov2019xgandalf} among the most widely used.  

PinkIndexer formulates indexing as a geometric constraint problem and was one of the first methods tailored to snapshot diffraction patterns (including pink-beam X-rays and electron diffraction) rather than conventional rotation-series data.
PinkIndexer assumes that approximate unit-cell parameters are known in advance.

XGANDALF introduced a modern formulation of indexing for serial crystallography based on an extended gradient-descent algorithm for lattice finding.  
It constructs a continuous score function from proximity to families of lattice planes and performs fast gradient-based refinement, making it well suited for real-time feedback and widely adopted in CrystFEL ~\cite{crystfel} pipelines.  
XGANDALF can operate both with a known unit cell and in a mode where the lattice parameters are inferred from the data, i.e., the unit cell is optional rather than strictly required.
However, improvements can be made in terms of robustness over several factors (noise, false peaks, and small number of reflections present).

More recently, TORO~\cite{gasparotto2023toro} introduced a PyTorch-based indexing framework that combines relaxed Laue score functions with robust optimization techniques, including least-trimmed-squares–style estimation and residual-threshold annealing, together with an speed efficient orientation–space exploration strategy, although memory heavy. Its advantage comes in the low number of iterations required for convergence.
TORO runs on GPUs, CPUs and other accelerators supported by PyTorch, making it an attractive candidate for real-time serial crystallography pipelines while matching or surpassing the accuracy of established indexers such as XGANDALF and MOSFLM on their tested benchmarks.
In the experiments reported, TORO is used with the unit-cell parameters provided to the indexer; the authors note that the method could in principle be extended to also guess an initial cell, but this is not explored in their benchmarks.

Despite these advances, neither XGANDALF nor TORO explicitly addresses issues arising from poorly conditioned reciprocal bases, nor do they exploit geometric tools such as reciprocal-basis reduction and Babai’s nearest-plane decoder. 
Their objectives are designed around relaxed Laue equations and fast online processing, which naturally prioritizes speed over small-$N$–specific robustness. 
These gaps motivate the development of indexing methods that remain accurate under small number of reflections present, noise, and basis distortion, and that can also maintain efficient memory usage if needed, as pursued in the present work, where we assume that approximate unit-cell parameters are known.

\subsection*{Our Contributions}
In this work we introduce a robust indexing algorithm for serial crystallography with the following main contributions:

\begin{itemize}
    \item We formulate indexing as a symmetry-aware lattice decoding problem and design a robust orientation-dependent loss that explicitly incorporates lattice symmetries and outlier trimming.
    \item We combine this loss with a reciprocal-space basis reduction and a nearest-plane integer decoder, leading to stable indexing even for skewed unit cells.
    \item We develop a dedicated small-$N$ regime, using a refined per-peak objective and local search, that substantially improves indexing rates for patterns with very few reflections.
    \item We show that the method can be implemented in a memory-efficient way on GPUs and evaluate it on three XFEL datasets, where it consistently matches or outperforms state-of-the-art indexers (TORO and XGANDALF), with particularly large gains in the small-$N$ regime.
\end{itemize}

This positions our method as a complementary tool to existing indexers: highly robust and memory-efficient, particularly effective for images with a small number of reflections, and for skewed unit cells.

%------------------------------------------------------------------------------
\section{Indexing Problem}
\label{sec:indexing-problem}

The goal of crystallographic indexing is to determine the orientation of a crystal and assign Miller indices to the observed diffraction peaks.  
In serial crystallography, each diffraction pattern contains a sparse set of reciprocal-space vectors
\[
    \mathcal{Q} = \{q_1, \ldots, q_N\} \subset \mathbb{R}^3,
\]
corresponding to detector-localized peak positions mapped into laboratory coordinates.  
Given the experimental geometry and approximate unit-cell parameters, the reciprocal lattice is described by a basis matrix
\[
    B \in \mathbb{R}^{3 \times 3}
\]
(typically chosen as the inverse transpose of a direct-space basis).  
For a rotation matrix \(R \in SO(3)\), the corresponding reciprocal lattice in the laboratory frame is generated by \(RB\).

Indexing can therefore be formulated as the problem of finding an orientation \(R\) such that each observed \(q_n\) is well explained by some Miller index \(h_n \in \mathbb{Z}^3\) through the relation
\[
    q_n \approx RB h_n.
\]
The assignment of \((h_n)\) is implicit: for a fixed \(R\), candidate indices are obtained by projecting each \(q_n\) onto the lattice spanned by \(RB\) and applying an integer-decoding rule (rounding, nearest-plane search, or similar).  
Thus, the core task is to identify the rotation \(R\) for which these decoded indices agree with the underlying crystal lattice structure.

It is important to note that the solution \((R, H)\) is not unique: the reciprocal lattice is invariant under the crystal’s symmetry group.  
If \(S\) is any symmetry operation, then \(RB h = R (S B) (S^{-1} h)\) generates the same set of reciprocal-lattice vectors.  
Therefore, indexing is defined only up to the symmetry group, and any valid orientation \(R\) must be interpreted modulo these symmetry-equivalent transformations.

\subsection{Indexing as an Optimization Problem}
\label{subsec:indexing-opt}

In mathematical optimization terms, the indexing task can be expressed as a joint inference problem over the crystal orientation and the corresponding Miller indices.  
Given observed reciprocal-space vectors
\(
    \mathcal{Q}=\{Q_1,\dots,Q_N\}\subset\mathbb{R}^3, \quad Q \in \mathbb{R}^{3 \times N}
\)
and a reciprocal basis \(B\) derived from the known unit-cell parameters, the ideal orientation \(R\in SO(3)\) and integer triples \(H_i\in\mathbb{Z}^3\) should satisfy
\[
    Q_i \approx R B H_i .
\]
for $Q_i$ the reciprocal-space vector of a true reflection.

A general formulation of the indexing problem is therefore
\begin{equation}
    (R^\star, H^\star)
    \;=\;
    \argmin_{R \in SO(3),\, H \in \mathbb{Z}^{3\times N}}
    L\!\left(Q,\, R B H\right),
    \label{eq:indexing-generic}
\end{equation}
where \(L\) is any suitable discrepancy measure between the observed peak
positions \(\mathcal{Q}\) and the lattice points predicted by the oriented
reciprocal basis \(R B H\).

or an analogous robust variant.  
Because crystallographic lattices are invariant under the Laue symmetry group \(\mathcal{G}\), solutions are not unique:  
\((R, H)\) and \((RS,\, S^{-1}H)\) represent the same physical orientation for any \(S\in\mathcal{G}\).

This abstract formulation highlights the intrinsic structure of the indexing problem: a highly non-convex search over rotations coupled with discrete integer assignments, defined only up to lattice symmetry.

%------------------------------------------------------------------------------
\section{Methods}

\subsection{Algorithm Overview}

Given reciprocal–space vectors $\{Q_i\}_{i=1}^N$, a reciprocal basis $B$,
and unit–cell parameters, the indexer seeks an orientation $R^\star\in SO(3)$
and Miller indices $\{H_i^\star\}_{i=1}^N$ such that
$Q_i \approx R^\star B H_i^\star$ for as many peaks as possible.
We precompute a reduced reciprocal basis and symmetry operators, define a
robust orientation–dependent loss, minimize this loss over $SO(3)$ using
differential evolution (DE), and finally decode the Miller indices using a nearest-lattice procedure based on Babai’s Nearest Plane Algorithm (Section~\ref{subsec:babai}) (with a local-search for patterns with small number of reflections)
The steps are summarized in Algorithm~\ref{alg:indexer}.

\begin{algorithm}[H]
  \caption{Robust symmetry–aware indexing (high-level)}
  \label{alg:indexer}
  \begin{algorithmic}[1]
    \Require Reciprocal vectors $\{Q_i\}_{i=1}^N$, basis $B$, trim fraction $\kappa$, small-$N$ threshold $N_{\mathrm{small}}$, DE population size $P$, mutation scale $F$, crossover probability $p_{\mathrm{cx}}$, maximum iterations $G_{\max}$, refinement checkpoint interval $T_{\mathrm{ref}}$

    \Ensure Orientation $R^\star\in SO(3)$, indices $\{H_i^\star\}_{i=1}^N\subset\mathbb Z^3$
    \State \textbf{Precomputation:}
    \Statex \quad Perform reciprocal basis reduction (see Section~\ref{subsec:reduction}) to obtain $B_{\mathrm{red}} = B\,U$ with $U\in SL(3,\mathbb Z)$
    \Statex \quad Compute $B_{\mathrm{red}} =Q_B R_B$ and symmetry operators for the lattice (in reciprocal space)
    \State Represent each candidate crystal orientation by a unit quaternion $q\in S^3$, with $R(q)\in SO(3)$ the corresponding rotation matrix
    \State Initialize the DE population by sampling quaternions uniformly on $S^3$ (see Section~\ref{subsec:de})

    \State \textbf{Define orientation loss} (\emph{see} Section~\ref{subsec:loss}):
    \If{$N \le N_{\mathrm{small}}$} \Comment{e.g.\ $N_{\mathrm{small}} = 25$}
      \State $L(q) \gets L_{\mathrm{small}}\big(R(q); Q, B_{\mathrm{red}}, Q_B, R_B, \text{symmetry}, \kappa\big)$
    \Else
      \State $L(q) \gets L_{\mathrm{big}}\big(R(q); Q, B_{\mathrm{red}}, Q_B, R_B, \text{symmetry}, \kappa\big)$
    \EndIf
    \State \textbf{Run DE with occasional local refinement} (\emph{see} Section~\ref{subsec:de} and ~\ref{subsec:refinement}):

    \While{DE not converged}
      \State Perform one DE generation on the quaternion population 
      \Statex \quad (mutation, crossover, normalization)
      \State Evaluate the loss $L(q)$ for all population members
        \If{generation index is a refinement checkpoint}
          \State $q_{\text{DE}} \gets \arg\min_q L(q)$ in the current population
          \State $R_{\text{DE}} \gets R(q_{\text{DE}})$
          \State Locally refine $R_{\text{DE}}$ (Babai/Local-Search + Procrustes) to obtain $R_{\text{ref}}$ (\emph{see} Section~\ref{subsec:babai})
          \State $R_{\text{best}} \gets \arg\min_{R \in \{R_{\text{best}},\,R_{\text{ref}}\}} L(R)$
        \EndIf

    \EndWhile
    \State \textbf{Select final orientation:}
    \State $R^\star(q^\star) \gets$ best orientation (argmin $L(q)$) among DE and shadow–refined candidates
    \State \textbf{Final indexing:}
    \For{$i = 1,\dots,N$}
      \State $y_i \gets Q_B^\top {R^\star}^\top Q_i$
      \State Decode reduced index $H_{\mathrm{red},i}$ by Babai / Local-search in the reduced basis (with symmetry)
      \State Map back to original indices: $H_i^\star \gets U\,H_{\mathrm{red},i}$
    \EndFor
    \State \Return $R^\star$, $\{H_i^\star\}_{i=1}^N$
  \end{algorithmic}
\end{algorithm}

\subsection{Differential Evolution}
\label{subsec:de}

Differential evolution (DE)~\cite{storn1997differential} is a population--based stochastic optimizer for real--valued, nonconvex objectives.
Given a loss $L(\theta)$ with $\theta\in\mathbb R^d$, DE maintains a population
\[
\Theta^{(g)} = \big\{\theta_1^{(g)},\dots,\theta_P^{(g)}\big\}
\]
at generation $g$, and iteratively produces a new population by three steps for each index $i\in\{1,\dots,P\}$:

\paragraph{Mutation.}
Select distinct indices $r_1, r_2 \neq i$ uniformly at random without replacement from the population (excluding $i$), and let $\theta_{\mathrm{best}}^{(g)}$ denote the current best individual.  
In the \texttt{best-1-bin} scheme we use, the mutant vector is
\[
v_i^{(g)}
= \theta_{\mathrm{best}}^{(g)}
  + F\bigl(\theta_{r_1}^{(g)} - \theta_{r_2}^{(g)}\bigr),
\qquad r_1, r_2 \neq i,\; r_1 \neq r_2.
\]

where $F>0$ is a mutation scale factor. The parameter $F$ controls the exploration step size in orientation space: larger values promote broader exploration, while smaller values bias the search toward local refinement.  
In our experiments we follow standard DE practice and sample $F$ uniformly in the interval $[0.5, 1]$ for each generation.

\paragraph{Crossover.}
A trial vector $u_i^{(g)}$ is formed by mixing coordinates from $v_i^{(g)}$ and $\theta_i^{(g)}$.
In binomial crossover, for each component $k\in\{1,\dots,d\}$,
\[
u_{i,k}^{(g)}
=
\begin{cases}
v_{i,k}^{(g)}, & \text{with probability } C_r,\\[0.3em]
\theta_{i,k}^{(g)}, & \text{otherwise},
\end{cases}
\]
with crossover rate $C_r\in[0,1]$. In practice we use the standard DE convention of setting $C_r$ to be $0.7$, which balances exploration and stability

\paragraph{Selection.}
The trial vector competes with the parent:
\[
\theta_i^{(g+1)}
=
\begin{cases}
u_i^{(g)}, & \text{if } L(u_i^{(g)}) \le L(\theta_i^{(g)}),\\[0.3em]
\theta_i^{(g)}, & \text{otherwise}.
\end{cases}
\]
Thus DE performs a stochastic, \emph{elitist} search over $\mathbb{R}^d$, in the sense that the current best solution is always preserved in the population, and is an efficient method for non-convex, non-smooth landscapes that can escape local minima.

\paragraph{Application to orientation search.}
In our indexer, the parameter $\theta$ is a unit quaternion $q\in S^3\subset\mathbb R^4$ representing an orientation $R(q)\in SO(3)$.
\\Note that unit quaternions provide a double cover of $SO(3)$, since $q$ and $-q$ represent the same rotation.  
Because our loss depends only on $R(q)$, this redundancy has no effect on the optimization.  
In practice, keeping both $q$ and $-q$ in the DE population is harmless: mutation, crossover, and selection operate directly on the quaternion parameters, so the two representatives typically follow different stochastic trajectories, providing additional exploration at negligible cost.

We apply DE to the quaternion coordinates using the loss
\[
L(q) \in \{L_{\mathrm{big}}(R(q)),\,L_{\mathrm{small}}(R(q))\}
\]
defined in Section~\ref{subsec:loss}, depending on the number of observed reflections, $N$.
After each mutation and crossover step we renormalize,
\[
q \leftarrow \frac{q}{\|q\|_2},
\]
to keep all population members on the unit 3--sphere.

DE is responsible for exploring the orientation space, while the Babai step provides accurate lattice decoding for each proposed orientation so that the loss can be evaluated reliably.

\paragraph{Retries and memory considerations.}
Indexing pipelines such as \textsc{CrystFEL} commonly rely on multiple retries with different seeds. DE is especially compatible with this practice, since different initializations or random seeds lead to distinct stochastic explorations of orientation space. Our implementation also maintains a very small memory footprint, as DE does not require large population size since it is more dependent on number of iterations.
In practice, DE does not require a large population, since increasing the number of generations often yields similar improvements in solution quality.  This well-known trade-off allows small populations to converge reliably in many settings, including ours, and keeps the memory footprint minimal. In practice, it is often recommended to set the population size between 5$d$-10$d$, where $d$ is the number of dimensions of candidates.

\subsection{Babai Rounding}
\label{subsec:babai}

Given a full--rank lattice basis $A\in\mathbb R^{3\times 3}$, the associated lattice is
\[
\Lambda(A) := \{A h : h \in \mathbb Z^3\}.
\]
The \emph{closest vector problem} (CVP) asks, for a point $x\in\mathbb R^3$, to find the lattice point $A h$ closest to $x$ in Euclidean distance.
Babai's nearest--plane algorithm~\cite{babai1986lovasz} is a classic and efficient approximation: it takes $x$ and produces an integer vector $\widehat h$ by rounding in a suitable orthonormal coordinate system.

\paragraph{Babai in QR coordinates.}
Let $A = Q R$ be a QR factorization with $Q\in O(3)$ orthogonal and $R$ upper triangular with $R_{ii}>0$.
For $x\in\mathbb R^3$ define
\[
y := Q^\top x.
\]
In the noiseless case $x = A h$ with $h\in\mathbb Z^3$ gives $y = R h$.
Babai's decoder computes an estimate $\widehat h\in\mathbb Z^3$ via backward substitution:
\[
\widehat h_i
= \operatorname{round}\!\Big(
   \frac{y_i - \sum_{j>i} R_{ij}\,\widehat h_j}{R_{ii}}
  \Big),
\qquad i = 3,2,1,
\]
and returns the lattice point $\widehat x := A\,\widehat h$.

We choose Babai rounding over the more common least-squares-plus-rounding approach (i.e., solving $\min_h \|Ah - x\|$ and then rounding $h$ to the nearest integer vector), as it offers clear advantages, as discussed in Section~\ref{subsec:babai-vs-round}.

\paragraph{Use in our indexer.}
In our setting $A = B_{\mathrm{red}} = B U$ is the reduced reciprocal basis (Section~\ref{subsec:reduction}), with QR factorization
\[
B_{\mathrm{red}} = Q_B R_B.
\]
For a fixed orientation $R\in SO(3)$ and peak $Q_i$, we work with
\[
y_i := Q_B^\top R^\top Q_i = R_B h_i + \varepsilon_i,
\]
where $h_i\in\mathbb Z^3$ are the (unknown) Miller indices in the reduced basis and $\varepsilon_i$ is noise.
Applying Babai's rule with $R_B$ yields an estimate
\[
\widehat H_{i}^{\mathrm{red}}(R) \in \mathbb Z^3,
\]
and we map back to the original reciprocal basis via $H_i(R) := U\,\widehat H_{i}^{\mathrm{red}}(R)$.
This Babai step is the core nearest--lattice decoder used in our objectives, with a local integer search around $\widehat H_{i}^{\mathrm{red}}(R)$ in the small--$N$ case (Section~\ref{subsec:loss}).

\subsection{Reciprocal Basis Reduction}
\label{subsec:reduction}

\paragraph{Introduction.}
The reciprocal lattice is
\[
\mathcal L^\ast=\{B\,h:\,h\in\mathbb Z^3\},
\]
where \(h\) are Miller indices and \(B\in\mathbb R^{3\times 3}\) is a reciprocal basis.
We replace \(B\) by a lattice–equivalent \emph{reduced} basis
\[
B_{\mathrm{red}} := B\,U,\qquad U\in SL(3,\mathbb Z),
\]
where \(SL(3,\mathbb Z)\) denotes the group of \(3\times 3\) integer matrices with determinant \(1\) (unimodular integer changes of basis).
Such a change preserves the lattice \(\mathcal L^\ast\) but re-labels Miller indices.
Reduced bases are chosen to avoid very short and highly correlated basis vectors: they are typically closer to orthogonal and better balanced in length, which improves the conditioning of lattice decoding.
As we show in Section~\ref{subsubsec:babai-basis}, in the reduced coordinates Babai’s nearest–plane decoder succeeds whenever the noise falls inside an axis–aligned box whose side lengths are given by the Gram–Schmidt lengths of the basis; enlarging these lengths, especially in the weakest directions, enlarges the success region and tends to increase the probability of correct indexing.

\paragraph{Index mapping, QR, and orientation.}
Unimodularity preserves the lattice and maps indices via
\[
H = U\,H_{\mathrm{red}},\qquad H_{\mathrm{red}}=U^{-1}H,
\]
so that indexing in the reduced basis (\(H_{\mathrm{red}}\)) is equivalent to indexing in the original basis (\(H\)).
Compute a QR decomposition of \(B_{\mathrm{red}}\):
\[
B_{\mathrm{red}} = Q_B R_B,\quad Q_B\in O(3),\ R_B\ \text{upper triangular with }(R_B)_{ii}>0,
\]
where \(O(3)\) is the group of orthogonal \(3\times 3\) matrices (rotations and reflections), and the diagonal entries \((R_B)_{ii}\) are the Gram–Schmidt lengths of the reduced basis vectors.
For a candidate orientation \(R\in SO(3)\) (the group of \(3\times 3\) rotation matrices), we rotate the measured reciprocal vectors \(Q_i\) into the reduced, orthonormal coordinates via
\[
y_i := Q_B^\top R^\top Q_i.
\]
In these coordinates the noiseless model is
\[
y_i = R_B\,H_{\mathrm{red},i},
\]
so each measurement lies exactly on the lattice generated by \(R_B\) when the Miller indices are correct.
Babai (nearest–plane) decoding then reduces to triangular back–substitution using \(R_B\).

\subsection{Objective Function For a Given Orientation}
\label{subsec:loss}

For a given orientation $R\in SO(3)$ the indexer should reward configurations where many observed peaks $\{Q_i\}_{i=1}^N$ lie close to reciprocal--lattice points of the form
\[
Q \approx R B H,\qquad H\in\mathbb Z^3,
\]
up to lattice symmetry.
We therefore define an orientation--dependent loss $L(R)$ that penalizes the distance from each rotated peak to its nearest symmetry--equivalent lattice point, and then aggregates these distances in a robust way.
The same construction is used for the quaternion parameterization $q$ via $R=R(q)$.

\paragraph{Symmetry--aware per--peak residual.}
Let $\{S_\ell\}_{\ell=1}^L$ denote the reciprocal--space rotation matrices representing the point--group (lattice) symmetries of the unit cell.
For a candidate orientation $R$ and peak $Q_i$, we first rotate into the crystal frame and then apply each symmetry:
\[
v_{i,\ell}(R) := S_\ell^\top R^\top Q_i.
\]
Expressing the residuals in the reduced QR coordinates proceeds in two steps.  
First, for each symmetry operator $S_\ell$ we define
\[
v_{i,\ell}(R) = S_\ell^\top R^\top Q_i .
\]
Next, projecting into the reduced coordinates gives
\[
y_{i,\ell}(R)
    := Q_B^\top v_{i,\ell}(R)
     = Q_B^\top S_\ell^\top R^\top Q_i
     = R_B H_{i,\ell}^\star + \varepsilon_{i,\ell}.
\]

for some (unknown) integer $H_{i,\ell}^\star\in\mathbb Z^3$ and noise $\varepsilon_{i,\ell}$.
We decode $H_{i,\ell}^\star$ by applying Babai's nearest--plane algorithm in the reduced basis:
\[
\widehat H_{i,\ell}^{\mathrm{red}}(R)
  := \mathrm{Babai}\big(R_B^{-1} y_{i,\ell}(R)\big)\ \text{(nearest–plane lattice decoding in the $R_B$ basis)}\in\mathbb Z^3,
\]
and map back to the original reciprocal basis via $H_{i,\ell}(R) := U\,\widehat H_{i,\ell}^{\mathrm{red}}(R)$.
The corresponding predicted reciprocal--space vector in the lab frame is
\[
\widehat Q_{i,\ell}(R) := R\,S_\ell\,B\,H_{i,\ell}(R).
\]
We define the symmetry--aware squared residual for peak $i$ as
\[
r_i^2(R)
:= \min_{l\in \{1, \dots ,L\}} \big\|Q_i - \widehat Q_{i,\ell}(R)\big\|_2^2,
\]
and write $r_i(R) = \sqrt{r_i^2(R)}$.
Intuitively, $r_i(R)$ is the best fit to peak $Q_i$ obtainable by applying any lattice symmetry followed by nearest--lattice decoding in the reduced basis.

\paragraph{Robust aggregation.}
Given $\{r_i(R)\}_{i=1}^N$ for a fixed $R$, we want a scalar loss $L(R)$ that is robust to false peaks and noise.
We use two closely related objectives, depending on the number of peaks $N$:
for $N>N_{\mathrm{small}}$ we use a trimmed mean of squared residuals with a self--calibrating scale estimate;
for $N\le N_{\mathrm{small}}$ we use a softened nearest--lattice objective that more carefully resolves individual peaks before trimming.
In both cases a trim fraction $\kappa\in(0,1]$ controls the fraction of peaks that we want to be explained by a given orientation. The higher the estimated fraction of false peaks, the lower the value of $\kappa$ should be.

\subsubsection{Large--$N$ loss $L_{\mathrm{big}}$}

For larger peak sets we rely on the fact that, for a good orientation $R$, a nontrivial fraction of peaks cluster near the lattice while the remainder are outliers.
We therefore aggregate the symmetry--aware residuals $\{r_i^2(R)\}_{i=1}^N$ using a two–stage robust procedure: trimming followed by winsorization.

\paragraph{Trimming and winsorization.}
Let $r_i^2(R)$ denote the squared residual for peak $i$ as defined above, and let
\[
h := \big\lceil \kappa N \big\rceil
\]
for a user--chosen trim fraction $\kappa\in(0,1]$.
We first select the $h$ peaks with the smallest squared residuals,
\[
\mathcal I_{\kappa}(R)
:= \arg\min_{I\subset\{1,\dots,N\},\,|I|=h}
      \sum_{i\in I} r_i^2(R),
\]
which corresponds to trimming away the worst $(1-\kappa)N$ peaks.
On this trimmed set $\mathcal I_{\kappa}(R)$ we compute an empirical upper cutoff as a high quantile of the retained residuals.
Specifically, letting
\[
\{s_j(R)\}_{j=1}^h 
:= \{r_i^2(R):\ i\in\mathcal I_{\kappa}(R)\}
\]
be the multiset of trimmed squared residuals, we define
\[
t^2(R)
:= \operatorname{Quantile}_{0.90}\big(\{s_j(R)\}_{j=1}^h\big),
\]
the empirical $90$th percentile of the $h$ smallest squared residuals.
We then winsorize the trimmed residuals by clipping any $s_j(R)$ that exceeds $t^2(R)$:
\[
\tilde s_j(R) := \min\{s_j(R),\,t^2(R)\},\qquad j=1,\dots,h.
\]

Intuitively, the trimming step enforces that any plausible orientation must explain at least a $\kappa$ fraction of peaks with relatively small residuals, while the winsorization step limits the influence of the worst points even among those retained as inliers.
This construction does not assume any specific parametric form for the residual distribution; it uses only order statistics.

\paragraph{Definition of $L_{\mathrm{big}}$.}
The large--$N$ loss is defined as the mean of the winsorized squared residuals on the trimmed set,
\[
L_{\mathrm{big}}(R)
:= \frac{1}{h} \sum_{j=1}^{h} \tilde s_j(R).
\]
Equivalently,
\[
L_{\mathrm{big}}(R)
= \frac{1}{|\mathcal I_{\kappa}(R)|}
  \sum_{i\in\mathcal I_{\kappa}(R)}
  \min\big\{r_i^2(R),\,t^2(R)\big\}.
\]

The additional winsorization makes the loss less sensitive to a small number of unusually large residuals even among the $\kappa N$ best peaks.
In practice this combination provides a high degree of robustness to false positives and noise when many reflections are available.

\subsubsection{Small--$N$ loss $L_{\mathrm{small}}$}

When the number of peaks $N$ is small (e.g.\ $N \le N_{\mathrm{small}}$), pure nearest--plane decoding can be sensitive:
a single rounding error on one reflection may flip its Miller indices, and with few peaks such misassignments are not averaged out (see Section~\ref{subsec:small_n_advantage}).
To mitigate this, we augment Babai with a small local search in index space around each candidate label before trimming.

\paragraph{Local integer neighborhood.}
For each orientation $R$ and symmetry operator $S_\ell$, Babai applied in the reduced basis produces a baseline label
\[
H_{i,\ell}^{\mathrm{Babai}}(R) \in \mathbb Z^3
\]
for peak $Q_i$.
We then consider a small cubic neighborhood of integer offsets
\[
\Delta \in \{-\delta,\dots,\delta\}^3 \subset \mathbb Z^3,
\]
where $\delta\in\mathbb N$ is a small radius (in our implementation, $\delta=1$ to account for rounding error, so there are $K=(2\delta+1)^3=27$ offsets).
For each offset $\Delta$, we form the candidate index
\[
H_{i,\ell,\Delta}(R) := H_{i,\ell}^{\mathrm{Babai}}(R) + \Delta
\]
and the corresponding predicted reciprocal--space vector in the lab frame
\[
\widehat Q_{i,\ell,\Delta}(R)
  := R\,S_\ell\,B\,H_{i,\ell,\Delta}(R).
\]
The associated squared residual is
\[
\rho_{i,\ell,\Delta}^2(R)
 := \big\|Q_i - \widehat Q_{i,\ell,\Delta}(R)\big\|_2^2.
\]

\paragraph{Per--peak residual for small $N$.}
For each peak $i$ we take the best fit over both symmetry and local integer offsets:
\[
r_i^2(R)
:= \min_{\ell\in\{1,\dots,L\}} \ \min_{\Delta\in\{-\delta,\dots,\delta\}^3}
   \rho_{i,\ell,\Delta}^2(R).
\]
Equivalently, $r_i^2(R)$ is the squared distance from $Q_i$ to the closest lattice point reachable by changing the Babai label by at most $\delta$ in each Miller index, after accounting for lattice symmetries.

\paragraph{Trimming and definition of $L_{\mathrm{small}}$.}
Given the per--peak residuals $\{r_i^2(R)\}_{i=1}^N$ for a fixed orientation, we apply the same least--trimmed--squares idea as in the large--$N$ case, but without additional winsorization.
Let
\[
h := \big\lceil \kappa N \big\rceil,
\]
and let $\mathcal I_{\kappa}^{\mathrm{small}}(R)$ be the indices of the $h$ smallest squared residuals:
\[
\mathcal I_{\kappa}^{\mathrm{small}}(R)
:= \arg\min_{I\subset\{1,\dots,N\},\,|I|=h}
      \sum_{i\in I} r_i^2(R).
\]
The small--$N$ loss is then defined as
\[
L_{\mathrm{small}}(R)
:= \frac{1}{h}
   \sum_{i\in\mathcal I_{\kappa}^{\mathrm{small}}(R)} r_i^2(R).
\]

Thus $L_{\mathrm{small}}$ is an LTS objective built on top of a more careful, locally refined per--peak residual:
for each reflection, we allow the Miller indices to move within a small cube around Babai’s guess and pick the best nearby lattice point.
This strictly enlarges the region in which a reflection is assigned its correct indices compared to pure Babai decoding, which is especially important when only a small number of peaks are available and individual rounding errors cannot be averaged out.

\subsection{Callback Function and Local Refinement}
\label{subsec:refinement}

The DE optimizer in Section~\ref{subsec:de} explores the orientation space globally, but its population is constrained to a finite step size and may approach a good basin only coarsely.
To improve accuracy once promising orientations have been found, we interleave DE with a local refinement step implemented as a callback: at prescribed generations, the current best orientation is extracted, locally refined using lattice information (Babai decoding and Procrustes alignment), and the refined orientation is injected back into the population if it improves the loss.

\paragraph{Inlier selection and relabeling.}
Given a candidate orientation $R$ and per--peak residuals $r_i^2(R)$ (Section~\ref{subsec:loss}), we first identify a set of inlier reflections $\mathcal I_\kappa(R)$ of size $h=\lceil \kappa N\rceil$ by keeping the $h$ smallest residuals (large--$N$ case) or using the corresponding small--$N$ trimmed set.
For each $i\in\mathcal I_\kappa(R)$ we then recompute Miller indices using Babai decoding (and local search when $N\le N_{\mathrm{small}}$):
\[
y_i := Q_B^\top R^\top Q_i,\qquad
\widehat H_{i}^{\mathrm{red}}(R) := \mathrm{Babai/Local}\big(R_B,y_i\big),\qquad
\widehat H_i(R) := U\,\widehat H_{i}^{\mathrm{red}}(R).
\]
The corresponding lattice predictions in the lab frame are
\[
\widehat Q_i(R) := B\,\widehat H_i(R).
\]

\paragraph{Orthogonal Procrustes refinement.}
Conditioned on the (re)estimated indices $\{\widehat H_i(R)\}_{i\in\mathcal I_\kappa(R)}$, we refine the orientation by solving an orthogonal Procrustes problem on the inlier set:
\[
R_{\mathrm{ref}}
:= \arg\min_{R\in SO(3)}
    \sum_{i\in\mathcal I_\kappa(R)} \big\|Q_i - R\,\widehat Q_i(R)\big\|_2^2.
\]
Let
\[
C := \sum_{i\in\mathcal I_\kappa(R)} Q_i\,\widehat Q_i(R)^\top \in\mathbb R^{3\times 3}
\]
be the $3\times 3$ cross--covariance matrix between observed and predicted reciprocal--space vectors.
Writing the singular value decomposition as $C = U_C \Sigma V_C^\top$, the unique Procrustes solution in $SO(3)$ is
\[
R_{\mathrm{ref}} = U_C\,D\,V_C^\top,\qquad
D := \operatorname{diag}(1,1,\det(U_C V_C^\top)).
\]
This $R_{\mathrm{ref}}$ is the least--squares optimal rotation mapping the predicted lattice vectors $\widehat Q_i(R)$ to the observed peaks $Q_i$ over the inlier set.

\paragraph{Callback logic and early stopping.}
In practice we expose this refinement as a callback from the DE optimizer:
at each refinement checkpoint we take the current best quaternion $q_{\mathrm{DE}}$, form $R_{\mathrm{DE}}=R(q_{\mathrm{DE}})$, compute the refined orientation $R_{\mathrm{ref}}$ as above, and, if $L(R_{\mathrm{ref}})<L(R_{\mathrm{DE}})$, replace the best individual in the population by $R_{\mathrm{ref}}$.
The optimization is stopped either when DE’s built–in convergence criterion on the population (controlled by its tolerance parameter) is satisfied, or when a stagnation criterion is met (the best loss has not improved by more than a small threshold over a fixed number of generations).

\subsection{Proofs and Theoretical Justification}

\subsubsection{Babai Advantages Over Least--Squares Rounding}
\label{subsec:babai-vs-round}

A very common heuristic for integer decoding is \emph{least--squares plus rounding}: given a basis $A$ and a point $x$, one solves the unconstrained least--squares problem
\[
  h_{\mathrm{LS}} := \arg\min_{h\in\mathbb R^n} \|Ah - x\|_2^2 = A^{+}x,
\]
and then rounds $h_{\mathrm{LS}}$ coordinatewise to the nearest integer vector.
This procedure can perform well when $A$ is close to orthogonal, but in general it admits no known dimension--only worst--case approximation guarantee for the closest vector problem (CVP): for sufficiently skew bases, the distance from the LS+round solution to the lattice can be made arbitrarily larger than the true CVP distance (see, e.g., the discussion in~\cite{agrell2002closest}).

Babai’s nearest--plane algorithm~\cite{babai1986lovasz}, applied in QR coordinates, \emph{does} come with such a guarantee.
For any lattice $\Lambda(A)$ in dimension $n$ and any target $t\in\mathbb R^n$, the Babai point $x_{\mathrm{Babai}}$ satisfies
\[
  \|x_{\mathrm{Babai}} - t\|_2
  \;\le\;
  \gamma_n\,\mathrm{dist}(t,\Lambda(A)),
\]
for some constant $\gamma_n$ depending only on $n$ and the quality of the basis.
For example, using an LLL--reduced basis one obtains a bound of the form $\gamma_n \le 2^{n/2}$, with significantly smaller factors available for stronger reductions~\cite{agrell2002closest}.
Since we always apply Babai in a reduced reciprocal basis $B_{\mathrm{red}}$ (Section~\ref{subsec:reduction}), our decoder inherits a bounded worst--case CVP error factor, whereas LS+rounding in the original reciprocal basis does not.

Additionally, if we model the noise in the QR coordinates by a Gaussian integer least--squares model
\[
  y = Ah + v,\qquad h\in\mathbb Z^n,\ v\sim\mathcal N(0,\sigma^2 I),
\]
then results from the box--constrained OILS (optimal integer least squares) literature (``box--constrained'' meaning the integer vector $h$ is assumed a priori to lie in a finite hyper-rectangular index set) give a complementary probabilistic comparison: when $h$ is (approximately) uniform over this bounded index set, the success probability of the box--constrained rounding detector is always no larger than that of the corresponding Babai detector for the same $A$ and $\sigma$~\cite{wen2013box}.
Under this simple stochastic model, Babai decoding is therefore at least as reliable as LS+rounding in terms of the probability of assigning the correct integer vector, in addition to enjoying the worst--case geometric bounds above.

\subsubsection{Basis Reduction Advantage with Babai}
\label{subsubsec:babai-basis}

\paragraph{Setup (reduced coordinates).}
Fix a unimodular $U\in SL(3,\mathbb Z)$ and set $B_{\mathrm{red}} := B\,U$.
Let $B_{\mathrm{red}} = Q_B R_B$ be a QR factorization with $Q_B\in\mathrm{O}(3)$ and $R_B$ upper–triangular with $(R_B)_{ii}>0$.
For a single measured reflection $Q_i$, we model it in the reduced basis as
\[
x = B_{\mathrm{red}} h + \eta,\qquad h\in\mathbb Z^3,
\]
where $h$ are the (unknown) Miller indices and $\eta$ is measurement noise in reciprocal space.
Premultiplying by $Q_B^\top$ gives
\[
y := Q_B^\top x \;=\; R_B h + \varepsilon,\qquad \varepsilon := Q_B^\top \eta. \quad\text{(noise in orthonormal QR coordinates)}
\]
In these coordinates, the indexing problem for one reflection is: given $y$, recover the correct integer vector $h$.

\paragraph{Decoder (Babai / nearest–plane).}
Babai's nearest–plane decoder in these coordinates is
\[
\widehat h_i^{\mathrm{Babai}}
=\operatorname{round}\!\Big(\frac{y_i-\sum_{j>i}(R_B)_{ij}\widehat h_j^{\mathrm{Babai}}}{(R_B)_{ii}}\Big),
\]
applied successively for $i=3,2,1$.
This is the standard lattice–decoding rule for an upper–triangular generator matrix.
Define the success set
\[
\mathcal S_{\mathrm{Babai}}
=\{\varepsilon:\ \widehat h^{\mathrm{Babai}}(y)=h\ \text{when } y=R_B h+\varepsilon\},
\]
and, for a given noise distribution of $\varepsilon$, the per–reflection success probability
\[
P_{\mathrm{succ}}(R_B)=\Pr(\varepsilon\in\mathcal S_{\mathrm{Babai}}).
\]
In crystallographic terms, $P_{\mathrm{succ}}(R_B)$ is the probability that a single noisy reciprocal–space peak is assigned the correct Miller indices when expressed in the reduced basis.

\paragraph{Theorem (Babai success region and Gram--Schmidt diagonals).}
Define the axis-aligned box
\[
\mathcal B(R_B)=\prod_{i=1}^3\big(-\tfrac12 (R_B)_{ii},\ \tfrac12 (R_B)_{ii}\big).
\]

\textbf{(A) Characterization of the success region.}
Babai’s decoder succeeds if and only if the noise coordinates satisfy
\[
|\varepsilon_i|<\tfrac12 (R_B)_{ii}\quad\text{for all } i=1,2,3.
\]
Equivalently,
\[
\mathcal S_{\mathrm{Babai}}=\mathcal B(R_B)
\quad\Rightarrow\quad
P_{\mathrm{succ}}(R_B)\;=\;\Pr(\varepsilon\in\mathcal B(R_B)).
\]

\emph{Intuition.}
Decoding proceeds from \(i=3\) up to \(1\).
Once the higher indices are decoded correctly, the \(i\)-th step rounds
\[
\frac{y_i-\sum_{j>i}(R_B)_{ij}h_j}{(R_B)_{ii}}
= h_i + \frac{\varepsilon_i}{(R_B)_{ii}}.
\]
If \(|\varepsilon_i| < (R_B)_{ii}/2\), the nearest integer is the true \(h_i\); conversely, if Babai returns the correct $h_i$, then necessarily \(|\varepsilon_i|<(R_B)_{ii}/2\).
Requiring this for all \(i\) yields the box $\mathcal B(R_B)$.

\textbf{(B) Enlarging the Gram--Schmidt diagonals increases $P_{\mathrm{succ}}$.}
Fix the noise distribution of $\varepsilon$ in the reduced coordinates.
Let $R_B$ and $R_B'$ be upper–triangular matrices with positive diagonals, and suppose
\[
(R_B')_{ii}\ \ge\ (R_B)_{ii}\quad\text{for all } i=1,2,3.
\]
Then Babai’s success probability cannot decrease:
\[
P_{\mathrm{succ}}(R_B')
= \Pr(\varepsilon\in\mathcal B(R_B'))
\;\ge\;
\Pr(\varepsilon\in\mathcal B(R_B))
= P_{\mathrm{succ}}(R_B),
\]
because enlarging each diagonal enlarges the box,
\[
\mathcal B(R_B)\subseteq\mathcal B(R_B'),
\]
and hence increases its measure under any fixed noise distribution for $\varepsilon$.

\paragraph{Relation to reciprocal--basis reduction.}
For a fixed lattice, unimodular changes of basis preserve the lattice volume,
\(\det B_{\mathrm{red}} = \prod_i (R_B)_{ii}\), so it is impossible to enlarge all Gram–Schmidt lengths simultaneously via basis changes alone.
In crystallographic practice, however, conventional reciprocal cells for high-symmetry systems (cubic, tetragonal, orthorhombic, hexagonal, trigonal) are already close to reduced; LLL/KZ reduction (~\cite{wen2017kz}) typically returns the same (or close to) basis up to permutations and sign changes, so \(P_{\mathrm{succ}}(R_B)\) is essentially unchanged in these cases.
For low-symmetry or skew cells (e.g., triclinic/monoclinic, or poorly chosen primitive bases), reciprocal–lattice reduction algorithms such as LLL or KZ are explicitly designed to remove nearly dependent and highly correlated basis vectors: they tend to increase the smallest Gram–Schmidt lengths and produce a more balanced, less skew basis while keeping the volume fixed.
Since $P_{\mathrm{succ}}(R_B)$ is coordinatewise non-decreasing in the diagonals \((R_B)_{ii}\), improving these ``weak directions'' enlarges the Babai success region and, under the continuous and moderately anisotropic noise distributions for $\varepsilon$ that arise from detector geometry and peak integration in serial crystallography, tends to maintain or improve the overall success probability of Babai decoding.
In particular, this translates into a higher chance of assigning correct Miller indices to noisy peaks, and provides a strong practical justification for applying reciprocal–basis reduction before Babai decoding in our indexing pipeline.

\subsubsection{Small--$N$ Local Search Justification and Advantage}
\label{subsec:small_n_advantage}

For a fixed orientation $R$ and basis $R_B$, let $Z_i\in\{0,1\}$ indicate whether Babai decoding assigns the correct Miller indices to reflection $i$, with $\Pr(Z_i=1)=p$ under an i.i.d.\ noise model in the QR coordinates.
The empirical fraction of correctly decoded reflections is
\[
\hat p_N := \frac{1}{N}\sum_{i=1}^N Z_i,\qquad
\mathbb E[\hat p_N]=p,\quad
\mathrm{Var}(\hat p_N)=\frac{p(1-p)}{N}.
\]
By the weak law of large numbers, $\hat p_N\to p$ in probability as $N\to\infty$, so for large $N$ the effect of individual Babai errors on the orientation score is likely averaged out after being handled by trimming.
When $N$ is small, however, $\mathrm{Var}(\hat p_N)$ is large and a single decoding error changes $\hat p_N$ by $1/N$ (e.g.\ $0.2$ for $N=5$), making the loss $L(R)$ much more sensitive to per-reflection rounding mistakes.

Our small--$N$ loss $L_{\mathrm{small}}$ mitigates this by augmenting Babai with a local integer neighborhood
$\Delta\in\{-\delta,\dots,\delta\}^3, \delta=1$ around the Babai label. Denote by $E_{\mathrm{Babai}}$ the event that pure Babai yields the correct $h_i$, and by $E_{\mathrm{local}}$ the event that the correct $h_i$ lies within this neighborhood and is selected by the local search.
By construction (and under the mild assumption that the correct lattice point
is the unique best fit within this neighborhood), we have
\[
E_{\mathrm{Babai}} \;\subseteq\; E_{\mathrm{local}},
\]
and therefore
\[
p' := \Pr(E_{\mathrm{local}}) \;\ge\; \Pr(E_{\mathrm{Babai}})=p.
\]

Thus the local search typically enlarges the per--reflection success region, increasing the probability that peaks are correctly indexed and stabilizing the orientation score in the small--$N$ regime.

%------------------------------------------------------------------------------
\section{Results}
\label{sec:results}
In this section we evaluate the proposed indexing algorithm on publicly available serial femtosecond crystallography datasets from the Coherent X-ray Imaging Data Bank (CXIDB)~\cite{maia2012cxidb}. 
We focus on protein systems collected at XFEL facilities, spanning different beamlines and detector geometries. 
For each dataset, all indexers are run using identical CrystFEL settings~\cite{crystfel}, with retries, and the unit-cell parameters are provided to every method, as our algorithm requires it. 
We enable the default validation checks in \texttt{indexamajig}.

\subsection{Datasets}
\label{subsec:datasets}
We benchmark our method on three CXIDB entries: 61, 62, and 83. 

\paragraph{CXIDB 61: POMGnT1 stem domain.}
Entry 61 contains a serial femtosecond crystallography dataset of the stem domain of human POMGnT1, collected at SACLA (beamline BL3) with a photon energy of \SI{13.0}{keV} (\SI{0.954}{\angstrom})~\cite{yamashita2017experimental}. 
The dataset was originally acquired for experimental phase determination using SeMet or Hg derivatization, and is distributed in CXIDB with the corresponding MPCCD detector geometry and associated PDB structures (5XFC, 5XFD, 5XFE).

\paragraph{CXIDB 62: ACG (Agrocybe cylindracea galectin).}
Entry 62 provides a second SFX dataset from the same study~\cite{yamashita2017experimental}, comprising diffraction images of ACG (Agrocybe cylindracea galectin) collected at SACLA BL3 under nominally identical beam conditions (\SI{13.0}{keV}, \SI{0.954}{\angstrom}). 
As for CXIDB~61, the data are supplied with facility and refined CrystFEL geometry files and share the same associated PDB entries (5XFC, 5XFD, 5XFE).

\paragraph{CXIDB 83: $\beta$-lactamase at megahertz repetition rate.}
Entry 83 corresponds to one of the datasets from the ``Megahertz serial crystallography'' experiment at the European XFEL~\cite{wiedorn2018megahertz}. 
It contains diffraction patterns of a complex of $\beta$-lactamase collected at the SPB/SFX beamline with a photon energy of \SI{9.30}{keV} (\SI{1.33}{\angstrom}) at megahertz repetition rates, and is distributed together with detector geometry and the associated PDB structures (6GTH, 6FTR). 
Compared to the SACLA datasets, CXIDB~83 features different detector layout, noise characteristics, and pulse structure, providing a complementary test case for robust indexing.

\subsection{Indexing rate}
\label{subsec:indexing-rate}

Table~\ref{tab:indexing-all} presents the overall indexing rate (in \%) for the three CXIDB datasets when using our method, TORO~\cite{gasparotto2023toro}, and XGANDALF~\cite{gevorkov2019xgandalf} within CrystFEL~\cite{crystfel}. 
Across all systems, our indexer gives the highest indexing rate: it is slightly better than XGANDALF on CXIDB~61, ahead by a few percent of both TORO and XGANDALF on CXIDB~62, and slightly better than TORO on CXIDB~83. \\
The main advantage/use case of our algorithm is shown in ~\ref{subsubsec:indexing-small-N}, where we focus on the small N patterns.

\begin{table}[H]  % requires \usepackage{float}
  \centering
  \caption{Indexing rate (in \%) on all diffraction patterns for the three CXIDB datasets. The total number of processed patterns is given in parentheses.}
  \label{tab:indexing-all}
  \begin{tabular}{lccc}
    \hline
    Dataset (patterns) & Ours & TORO & XGANDALF \\
    \hline
    CXIDB 61 (35\,294) & \textbf{90.6} & 88.2 & 89.7 \\
    CXIDB 62 (28\,835) & \textbf{95.2} & 93.0 & 92.6 \\
    CXIDB 83 (17\,043) & \textbf{91.5} & 90.6 & 85.3 \\
    \hline
  \end{tabular}
\end{table}

A detailed step-by-step description of the experimental procedure to obtain indexing rates, including the parameters used in CrystFEL, will be provided in the Appendix

% Figures~\ref{fig:indexing-vs-N-61}--\ref{fig:indexing-vs-N-83} show the indexing rate as a function of the number of detected reflections $N$ for each dataset. 
% The curves indicate that all three methods perform similarly once a large number of reflections is available, while our method maintains a clear advantage when $N$ is small.

\subsubsection{Indexing rate for $N \le 25$}
\label{subsubsec:indexing-small-N}

To highlight performance in the challenging small-$N$ regime, we restrict the analysis to patterns with at most $25$ detected reflections. 
Table~\ref{tab:indexing-smallN} summarizes the indexing rate on this subset. 
With $N \le 25$, our method indexes substantially more patterns than TORO and XGANDALF on all three datasets.  
In contrast to the baselines, which suffer a marked drop in indexing rate for sparse patterns, our indexer remains close to its full-dataset performance, underscoring the effectiveness of the small-$N$ loss and local refinement.

\begin{table}[H]  % requires \usepackage{float}
  \centering
  \caption{Indexing rate (in \%) restricted to patterns with at most $25$ reflections ($N \le 25$). The number of such patterns is given in parentheses.}
  \label{tab:indexing-smallN}
  \begin{tabular}{lccc}
    \hline
    Dataset ($N \le 25$ patterns) & Ours & TORO & XGANDALF \\
    \hline
    CXIDB 61 (2\,657) & \textbf{97.2} & 79.6 & 85.9 \\
    CXIDB 62 (3\,881) & \textbf{95.9} & 74.2 & 72.3 \\
    CXIDB 83 (2\,059) & \textbf{89.2} & 76.3 & 83.2 \\
    \hline
  \end{tabular}
\end{table}

\subsubsection{Indexing Rate vs.\ Number of Reflections}
\label{subsubsec:indexing-vs-n}

\begin{figure}[H]
    \centering

    \begin{subfigure}{0.32\linewidth}
        \centering
        \includegraphics[width=\linewidth]{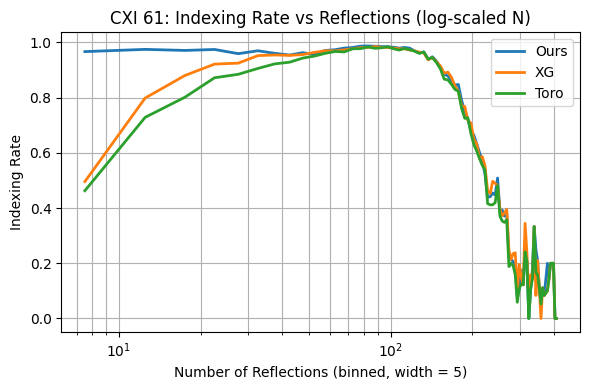}
        \caption{CXIDB~61}
        \label{fig:indexing-vs-N-61}
    \end{subfigure}
    \hfill
    \begin{subfigure}{0.32\linewidth}
        \centering
        \includegraphics[width=\linewidth]{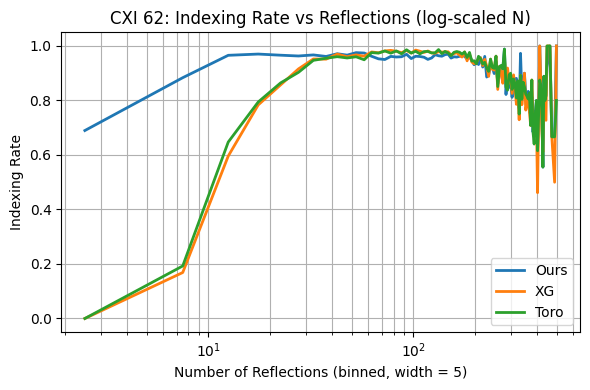}
        \caption{CXIDB~62}
        \label{fig:indexing-vs-N-62}
    \end{subfigure}
    \hfill
    \begin{subfigure}{0.32\linewidth}
        \centering
        \includegraphics[width=\linewidth]{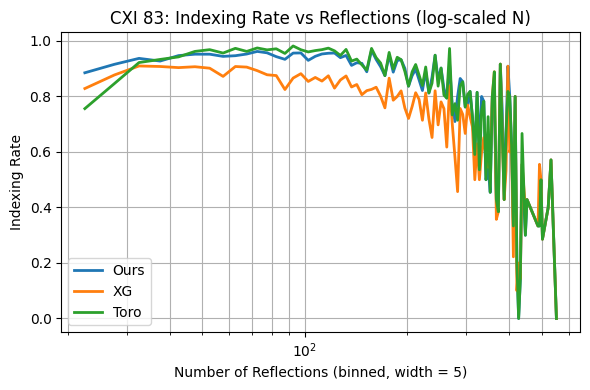}
        \caption{CXIDB~83}
        \label{fig:indexing-vs-N-83}
    \end{subfigure}

    \caption{Indexing rate vs.\ number of reflections for CXIDB~61, 62, and 83.}
    \label{fig:indexing-vs-N-all}
\end{figure}

Figures~\ref{fig:indexing-vs-N-61}--\ref{fig:indexing-vs-N-83} show the indexing rate as a function of the number of detected reflections $N$ for CXIDB~61, 62, and 83.  
Across all datasets, the three indexers converge to similar performance once a sufficiently large number of reflections is available, except for CXIDB 83m where XGandalf underperforms across most $N$.  
However, in the sparse-reflection regime (small $N$), our method consistently maintains a clear advantage: it yields higher indexing rates for small $N$ on all datasets, and remains competitive for larger $N$
These results reinforce the trends observed in Table~\ref{tab:indexing-smallN}, confirming that the proposed small-$N$ loss and refinement steps provide a decisive benefit when only a few reflections are present, while remaining competitive for well-populated patterns.

\subsection{Runtime and Memory Usage}
\label{subsec:runtime}

To assess computational efficiency, we measured the wall-clock time and GPU memory consumption when indexing batches of diffraction patterns.
Runtime was recorded using Python's \texttt{time} module, while GPU memory was measured through PyTorch's built-in memory profiling utilities.
All GPU experiments were performed on the same NVIDIA A100 under identical software and memory constraints, ensuring that the comparison between indexers reflects algorithmic differences rather than hardware variations.
Tables~\ref{tab:runtime-n33} and \ref{tab:runtime-n15} report the runtime per batch (with the batch size indicated) together with the peak GPU memory allocated and reserved during execution.
Additional implementation details on the measurement setup will be provided in Appendix.

\begin{table}[H]
    \centering
    \caption{Runtime per batch and peak GPU memory usage for 100 random CXI62 patterns with $N=33$ reflections.}
    \label{tab:runtime-n33}
    \begin{tabular}{@{}lcccc@{}}
        \hline
        \textbf{HW} & \textbf{Indexer} & \textbf{Batch} & \textbf{Time / batch (s)} & \textbf{GPU mem. (alloc/res) [GB]} \\
        \hline
        AMD EPYC (CPU)   & TORO & 1   & 0.37 $\pm$ 0.05 & N/A \\
        AMD EPYC (CPU)   & Ours & 1   & 1.19 $\pm$ 0.57 & N/A \\
        NVIDIA A100 (GPU)& TORO & 100 & 0.56 $\pm$ 0.09 & 11.69 / 22.51 \\
        NVIDIA A100 (GPU)& Ours & 100 & 7.06 $\pm$ 0.09 & 0.41 / 0.55 \\
        \hline
    \end{tabular}
\end{table}

\begin{table}[H]
    \centering
    \caption{Runtime per batch and peak GPU memory usage for 100 random CXI62 patterns with $N=15$ reflections.}
    \label{tab:runtime-n15}
    \begin{tabular}{@{}lcccc@{}}
        \hline
        \textbf{HW} & \textbf{Indexer} & \textbf{Batch} & \textbf{Time / batch (s)} & \textbf{GPU mem. (alloc/res) [GB]} \\
        \hline
        AMD EPYC (CPU)   & TORO & 1   & 0.26 $\pm$ 0.05 & N/A \\
        AMD EPYC (CPU)   & Ours & 1   & 3.51 $\pm$ 1.64 & N/A \\
        NVIDIA A100 (GPU)& TORO & 100 & 0.16 $\pm$ 0.03 & 5.34 / 10.26 \\
        NVIDIA A100 (GPU)& Ours & 100 & 10.34 $\pm$ 0.03 & 1.21 / 1.34 \\
        \hline
    \end{tabular}
\end{table}
Across GPU experiments, our indexer is markedly more memory-efficient: for identical batch sizes, it requires orders of magnitude less GPU memory than TORO.  
However, in terms of raw speed, TORO is consistently much faster, typically by at least one order of magnitude on both CPU and GPU.  
This difference is expected and arises from several characteristics of our optimization procedure:

\begin{enumerate}
    \item \textbf{Iteration-heavy optimization.}  
    Differential Evolution (DE) requires many population updates, each involving several function evaluations.  
    Consequently, even modest per-iteration cost accumulates substantially over the optimization trajectory.

    \item \textbf{Exploration-first orientation search.}  
    We use DE to explore the orientation space.  
    While DE provides strong robustness to noise, helps escape local minima, and has low memory requirements on GPU, its crossover--mutation cycle introduces non-negligible computational overhead.  
    These operations are inexpensive individually, but the cumulative effect over many iterations leads to higher runtime.

    \item \textbf{Additional local search for small-$N$ patterns.}  
    For patterns with $N \leq 25$, we apply a small local-search step.  
    As established in Section~\ref{subsec:small_n_advantage}, this refinement is critical for small-$N$ robustness but introduces extra computational cost.  
    The increase in runtime is justified: Table~\ref{tab:indexing-smallN} and Figures~\ref{fig:indexing-vs-N-61}--\ref{fig:indexing-vs-N-83} show that our method dramatically outperforms other indexers in the low-reflection regime, where noise and ambiguity are typically highest.
\end{enumerate}

\noindent
\textbf{Notably}, because DE is stochastic, it also benefits from multiple retries: rerunning the population with different random initializations can lead to a total different exploration strategy of the space, which is particularly valuable for noisy datasets with narrow valleys.

\noindent
From a deployment perspective, current real-time pipelines at XFEL facilities typically process diffraction patterns sequentially with tight per-pattern latency constraints.  
In this online setting, our present implementation which relies on DE iterations and an additional refinement step for small-$N$ patterns is better suited to offline reprocessing than to live feedback, especially in memory-constrained environments, and when the number of small-$N$ images is dominant and noisy.
\textbf{Nevertheless}, because most of the computational overhead comes from DE's exploration of the orientation space and the need for many iterations to converge, a promising direction for future work, particularly if one wishes to prioritize speed over memory usage such as in online deployment, is to replace our Differential Evolution–based exploration of the orientation space with a faster sampling strategy inspired by TORO~\cite{gasparotto2023toro}, while retaining the objective functions introduced in this work.

Overall, although TORO achieves superior runtime on both CPUs and GPUs, our method provides competitive throughput on modern GPUs and maintains extremely low memory usage due to the population-based nature of DE, which trades memory for iterative refinement.  
Given the significantly higher indexing rates we obtain, especially for small-$N$ patterns, the additional runtime is often justified and acceptable for offline processing.  
In addition, the much smaller memory footprint means that, on GPUs, substantially larger batch sizes can be used on the same hardware, which can narrow the effective gap in frames processed per second compared to TORO.  
Similarly, on multi-core CPUs, the lower per-process memory usage allows more workers to be run in parallel before exhausting system memory, further improving throughput.

%------------------------------------------------------------------------------
\section{Discussion and Future Work}
\label{sec:discussion}

\subsection{Advantages}
Our approach offers several key strengths.  
First, the use of Differential Evolution (DE), combined with our objective functions and Babai-based evaluation, yields strong robustness to noise and maintains high accuracy even in the small-$N$ regime (with the help of the local-search), where most indexers degrade.  

Second, using Babai’s nearest-plane algorithm instead of least-squares rounding can improve the reliability of Miller index assignment, especially for skewed or ill-conditioned unit cells.  
This is further reinforced by reciprocal-space basis reduction, which produces a more orthogonal basis and improves the success probability of Babai’s nearest-plane algorithm as seen in ~\ref{subsec:babai-vs-round}.

Finally, because DE is population-based and does not rely on dense orientation grids or large precomputed structures, the method is extremely memory-efficient on GPUs, often an order of magnitude lighter than TORO for identical batch sizes. This low memory usage can be traded for higher parallelism, either through larger GPU batch sizes or more concurrent CPU workers, which helps tighten the effective speed gap relative to TORO despite higher per-batch runtime.

These properties make the algorithm particularly attractive for indexing of noisy datasets and scenarios dominated by low-$N$ patterns, especially since DE can benefit from multiple stochastic retries.

\subsection{Limitations}
The main limitation of the current implementation is runtime.  
DE requires many generations and repeated objective evaluations to sufficiently explore the orientation space, and the additional refinement step used for small-$N$ patterns further increases the computational cost.  
As a result, the method is considerably slower than TORO and XGANDALF on both CPUs and GPUs and is not suited for online deployment, where low latency per diffraction pattern is essential.  

\subsection{Applications and Integration}
Given its accuracy and low memory requirements, the method is well aligned with offline reprocessing pipelines, where full datasets are available on disk and batching can be exploited to achieve high throughput.  
In contrast, real-time pipelines process images sequentially and require rapid per-pattern indexing, making our current DE-based approach less suitable for live feedback.

\paragraph{Future Work. }A promising direction for future work is to accelerate orientation and make it suitable for online pipelines by replacing DE with a more speed-efficient sampling strategy, such as the basis-orientation search used in TORO~\cite{gasparotto2023toro}, while preserving the robustness advantages of our objective functions.  
Such a hybrid design could potentially bridge the gap between the small-$N$ performance of our method and the low-latency requirements of online processing, enabling deployment in both offline and live-crystallography contexts.

%------------------------------------------------------------------------------
\section{Conclusion}

We introduced a new crystallographic indexing algorithm that combines Differential Evolution–based orientation search, a symmetry-aware objective function, Babai-based integer assignment, and reciprocal-space basis reduction. Our method assumes known/estimated unit-cell parameters.
Across three XFEL datasets, the method achieves consistently higher indexing rates than existing indexers, particularly for small-$N$ patterns where methods such as TORO and XGANDALF experience significant degradation.  
This robustness arises from the combination of our symmetry-aware loss, the use of Babai’s nearest-plane algorithm in a reduced reciprocal basis, and a dedicated local search applied when only a few reflections are available.  

Although the algorithm is slower than existing methods due to the iterative nature of DE, it remains highly memory-efficient on modern GPUs and is well suited for offline reprocessing, particularly for noisy datasets and cases dominated by sparse diffraction patterns.  
The improved accuracy, especially for small $N$, can offset the additional runtime in practical scenarios.
Moreover, the markedly lower memory footprint means that larger GPU batches and more parallel CPU workers can be employed on the same hardware, so that, in practice, the effective indexing throughput can approach that of but more memory-intensive methods such as TORO.

Future work involves making the indexer more suitable for online pipelines by replacing DE with a faster orientation-sampling strategy, similar to the one employed in TORO, offering a promising direction for reducing runtime while preserving our advantages in our objective functions and reductions.  
Such approach could offer variants of our algorithm for both offline and online pipelines, bridging the gap between high accuracy at sparse and noisy reflection patterns and the low-latency requirements of real-time serial crystallography.

%------------------------------------------------------------------------------
% \appendix

% \section{Appendix: Dataset and experiments procedure}
% \label{appendix:indexing-details}

% \section{Runtime Measurement Details}
% \label{appendix:runtime-details}

%------------------------------------------------------------------------------
\begin{acknowledgements}
This research used resources of the National Energy Research Scientific Computing Center (NERSC), a Department of Energy User Facility using NERSC award [ABC]-ERCAP [12345].
This work used the resources of the SLAC Shared Science Data Facility (S3DF) at SLAC National Accelerator Laboratory. S3DF is a shared High-Performance Computing facility, operated by SLAC, that supports the scientific and data-intensive computing needs of all experimental facilities and programs of the SLAC National Accelerator Laboratory. SLAC is operated by Stanford University for the U.S. Department of Energy’s Office of Science.
\end{acknowledgements}

\begin{funding}
This work was supported by the U.S. Department of Energy, Office of Science, Basic Energy Sciences under Contract No. DEAC02-76SF00515. 
\end{funding}

\ConflictsOfInterest{The authors declare no conflicts of interest.}

\DataAvailability{
The datasets used in this study are publicly available through the Coherent X-ray Imaging Data Bank (CXIDB). 
Code and implementation details will be released at a later date.
}

\bibliography{iucr}

@article{storn1997differential,
  author  = {Storn, Rainer and Price, Kenneth},
  title   = {Differential Evolution -- A Simple and Efficient Heuristic for Global Optimization over Continuous Spaces},
  journal = {Journal of Global Optimization},
  year    = {1997},
  volume  = {11},
  number  = {4},
  doi     = {10.1023/A:1008202821328}
}

@article{babai1986lovasz,
  author  = {Babai, L{\'a}szl{\'o}},
  title   = {On Lovasz' Lattice Reduction and the Nearest Lattice Point Problem},
  journal = {Combinatorica},
  year    = {1986},
  volume  = {6},
  number  = {1},
}

@article{agrell2002closest,
  author  = {Agrell, Erik and Eriksson, Thomas and Vardy, Alexander and Zeger, Kenneth},
  title   = {Closest Point Search in Lattices},
  journal = {IEEE Transactions on Information Theory},
  year    = {2002},
  volume  = {48},
  number  = {8},
  doi     = {10.1109/TIT.2002.800499}
}

@article{wen2013box,  
  author  = {Wen, Zaiwen and Yin, Wotao},
  title   = {A Feasible Method for Optimization with Orthogonality Constraints},
  journal = {Mathematical Programming},
  year    = {2013},
  volume  = {142},
  number  = {1--2},
  doi     = {10.1007/s10107-012-0584-1}
}

@article{maia2012cxidb,
  author  = {Maia, Filipe R. N. C.},
  title   = {The Coherent X-ray Imaging Data Bank},
  journal = {Nature Methods},
  year    = {2012},
  volume  = {9},
  number  = {9},
  doi     = {10.1038/nmeth.2110}
}

@article{crystfel,
  author  = {White, Thomas A. and Mariani, Valerio and Brehm, Wolfgang and Yefanov, Oleksandr and Barty, Anton and Beyerlein, Kenneth R. and Chervinskii, Fedor and Galli, Lorenzo and Gati, Cornelius and Nakane, Takanori and Tolstikova, Alexandra and Yamashita, Keitaro and Yoon, Chun Hong and Diederichs, Kay and Chapman, Henry N.},
  title   = {Recent Developments in CrystFEL},
  journal = {Journal of Applied Crystallography},
  year    = {2016},
  volume  = {49},
  number  = {2},
  doi     = {10.1107/S1600576716004751}
}

@article{yamashita2017experimental,
  author  = {Yamashita, Keitaro and Kuwabara, Naoyuki and Nakane, Takanori and Murai, Tomohiro and Mizohata, Eiichi and Sugahara, Michihito and Takegawa, Shun-ichi and Ohsaki, Yuri and Takagi, Kenji and Noguchi, Hiroshi and Masuda, Tomoko and Kato, Kenji and Nango, Eriko and Tono, Kensuke and Joti, Yasumasa and Kameshima, Tetsuo and Yabashi, Makina and Yamamoto, Masaki},
  title   = {Experimental Phase Determination with Selenomethionine or Mercury-Derivatization in Serial Femtosecond Crystallography},
  journal = {IUCrJ},
  year    = {2017},
  volume  = {4},
  number  = {5},
  doi     = {10.1107/S2052252517008557}
}

@article{wiedorn2018megahertz,
  author  = {Wiedorn, Max O. and Oberth{\"u}r, Dominik and Bean, Richard and Schubert, Robin and Werner, Nadine and Abbey, Brian and Aepfelbacher, Andreas and others},
  title   = {Megahertz Serial Crystallography},
  journal = {Nature Communications},
  year    = {2018},
  volume  = {9},
  number  = {1},
  doi     = {10.1038/s41467-018-06156-7}
}

@article{gasparotto2023toro,
  author  = {Gasparotto, Piero and Barba, Luis and Stadler, Hans-Christian and Assmann, Greta and Mendonça, Henrique and Ashton, Alun W. and Janousch, Markus and Leonarski, Filip and Béjar, Benjamín},
  title   = {TORO Indexer: A PyTorch-based indexing algorithm for kilohertz serial crystallography},
  journal = {Journal of Applied Crystallography},
  year    = {2024},
  volume  = {57},
  number  = {4},
  doi     = {10.1107/S1600576724003182}
}

@article{gevorkov2019xgandalf,
  author  = {Gevorkov, Yaroslav and Yefanov, Oleksandr and Barty, Anton and White, Thomas A. and Tolstikova, Alexandra and Wiedorn, Max O. and Meents, Alke and Grigat, Ralf-Rainer and Chapman, Henry N. and Brehm, Wolfgang},
  title   = {XGANDALF -- Extended Gradient Descent Algorithm for Lattice Finding},
  journal = {Acta Crystallographica Section A: Foundations and Advances},
  year    = {2019},
  volume  = {75},
  number  = {5},
  doi     = {10.1107/S2053273319010593}
}

@article{wen2017kz,
  author = {Wen, Jinming and Chang, Xiao-Wen},
  title  = {On the KZ Reduction},
  journal= {arXiv preprint arXiv:1702.08152},
  year   = {2017},
  url    = {https://arxiv.org/abs/1702.08152}
}

@article{gevorkov2020pinkindexer,
  author  = {Gevorkov, Yaroslav and Barty, Anton and Mariani, Valerio and Yefanov, Oleksandr and Brehm, Wolfgang and Tolstikova, Alexandra and Chervinskii, Fedor and Chapman, Henry N. and White, Thomas A.},
  title   = {pinkIndexer -- a universal indexer for pink-beam X-ray and electron diffraction snapshots},
  journal = {Acta Crystallographica Section A: Foundations and Advances},
  year    = {2020},
  volume  = {76},
  number  = {2},
  pages   = {121--131},
  doi     = {10.1107/S2053273319015559}
}

\end{document}